# Trilayer TMDC Heterostructures for MOSFETs and Nanobiosensors

Kanak Datta, Abir Shadman, Ehsanur Rahman, Quazi D. M. Khosru

Department of Electrical and Electronic Engineering
Bangladesh University of Engineering and Technology
Dhaka, Bangladesh
* Email: kanakeee08@gmail.com

*Abstract*— Two dimensional materials such as Transition Metal Dichalcogenides (TMDC) and their bi-layer/tri-layer heterostructures have become the focus of intense research and investigation in recent years due to their promising applications in electronics and optoelectronics. In this work, we have explored device level performance of trilayer TMDC heterostructure ($MoS_2/MX_2/MoS_2$; M=Mo or, W and X=S or, Se) Metal Oxide Semiconductor Field Effect Transistors (MOSFETs) in the quantum ballistic regime. Our simulation shows that device 'on' current can be improved by inserting a $WS_2$ monolayer between two $MoS_2$ monolayers. Application of biaxial tensile strain reveals a reduction in drain current which can be attributed to the lowering of carrier effective mass with increased tensile strain. In addition, it is found that gate underlap geometry improves electrostatic device performance by improving sub-threshold swing. However, increase in channel resistance reduces drain current. Besides exploring the prospect of these materials in device performance, novel trilayer TMDC heterostructure double gate Field Effect Transistors (FETs) are proposed for sensing Nano biomolecules as well as for pH sensing. Bottom gate operation ensures these FETs operating beyond Nernst limit of 59 mV/pH. Simulation results found in this work reveal that scaling of bottom gate oxide results in better sensitivity while top oxide scaling exhibits an opposite trend. It is also found that, for identical operating conditions, proposed TMDC FET pH sensors show super-Nernst sensitivity indicating these materials as potential candidates in implementing such sensor. Besides pH sensing, all these materials show high sensitivity in the sub-threshold region as a channel material in nanobiosensor while $MoS_2/WS_2/MoS_2$ FET shows the least sensitivity among them.

◉ *Introduction:*

Layered van der Waals materials, such as metal dichalcogenides, few layers thick or exfoliated down to a single layer, have become the subject of extensive research in recent times [1][2]. *Ab-initio* simulation of electronic structures of monolayer Transition Metal Dichalcogenides (TMDC) materials reveals tunability in bandgap and electronic effective mass at conduction band minima under biaxial strain application [3][4]. Stacking multiple layers of on top of each other of these materials also leads to interesting changes in electronic properties [5][6][7]. The presence of intrinsic bandgap of monolayer and multilayer two dimensional (2D) materials, tunability of electronic properties with layer thickness, lower rate of electronic mobility degradation with dimensional scaling, and scalability down to monolayer dimension makes these materials suitable for electronic device application. Although $MoS_2$ is the most widely studied and investigated [8][9][10] TMDC material, high-performance Metal Oxide Semiconductor Field Effect Transistors (MOSFETs) have been implemented with other TMDC materials as well [11][12]. Recently, several studies have been performed on the modeling and projection of 2D Field Effect Transistors (FETs) for sub-10 nm Very-Large-Scale-Integration (VLSI) applications, further emphasizing their potential for ultra-scale high-performance electronic devices [13][14][15]. Besides, electronic and optoelectronic devices based on 2D bilayer heterostructures have been studied and investigated [16][17][18]. Amazing improvement in fabrication and processing technology over the last few years have allowed possible growth of these multilayer 2D heterostructures [19][20][21]. In addition to bilayer 2D heterostructures, trilayer TMDC heterostructures based on $MoS_2$ have also been studied using first principle simulations [22].

Ion-sensitive field-effect transistor (ISFET), also known as pH sensor, has been widely used to measure ion's concentrations ($H^+$ or $OH^-$) in a solution. The pH sensitivity (mV/pH) for a conventional single-gated ISFET, which is defined by the change of threshold voltage ($V_T$) at a given amount of pH change, is 59mv/pH and defines the Nernst limit. The Nernst limit of sensitivity (59 mV/pH) in single-gated ISFET can be overcome through the use of double gated field effect transistors as reported in the literature [23][24]. Besides pH sensing, ISFETs have also been modified to detect biomolecules like Deoxyribonucleic acid (DNA), Protein and biomarkers indicative of various diseases. In biosensing application, the dielectric layer is functionalized with specific receptors for selectively capturing the desired target biomolecules. The charged biomolecules create similar effects as created by site binding surface charge in the oxide-electrolyte interface in a normal pH sensor. Except for few recently reported experimental works on $MoS_2$ sensors [25][26][27][28] most of the works until now of super-Nernst pH and Nano biosensor involve silicon on insulator technology. However, no work, either simulation or experimental has been reported on trilayer TMDC heterostructure FET as potentiometric biosensor yet. The emerging 2D graphene sheets are evaluated for their superior sensing capacity [29] because of high surface to volume ratio. However, unlike mono or multilayer $MoS_2$ having a finite bandgap, graphene lacks bandgap which results in a large leakage current. Few-layer $MoS_2$ FETs are being considered as an attractive alternative to current sensor technology since their transport characteristics is extremely responsive to external stimulations[30].

In this work, we have explored the application of $MoS_2$ based van der Waal trilayer heterostructures as channel materials in sub-10 nm MOSFET and Double Gate Field Effect Transistor (DGFET) sensor applications. We have performed first-principle study on the electronic properties of trilayer materials using open source simulation framework Quantum Espresso [31]. The effect of bi-axial strain on electronic structure and properties was also observed using the same simulation package. Using parameters obtained from first principle simulation, a 10 nm double gate MOSFET is simulated using effective mass Hamiltonian approach in non-equilibrium Green's function (NEGF) formalism [32][33]. The effect of interband tunneling has been taken into account to get a better view of the sub-threshold operation of the device. Besides investigating the application of TMDC materials in high-performance electronic device, we have also proposed an application of these materials in potentiometric biosensing. A comparative study of their sensitivity dependence on various physical parameters like top and bottom oxide thickness as well as device operation regime is carried out to maximize their detection capability in pH and biosensor application. For pH and biomolecule sensing, Schrodinger equation coupled with nonlinear Poisson equation, which incorporates Boltzmann distribution in the electrolyte region, is solved self-consistently to calculate spatial charge and electrostatic potential distributions within the device. We have avoided simplification like Debye-Hückel approximation in an attempt to provide an accurate result. At the same time, the quantum mechanical charge density in the semiconductor have been taken into account. Finally, a drift-diffusion current model is used to measure the sensitivity of these devices.

◙ *Simulation Procedure:*

### A. Electronic Structure Simulation:

In this work, we have used $MoS_2/MX_2/MoS_2$ (M=W or, Mo; X= S or, Se) trilayer heterostructures as channel materials. We report the simulation results obtained using AAA stacking configuration only. The AAA stacking configuration of the trilayer material is shown in Fig. 1(d). For the electronic structure simulation, we have used pw.x package of the open source simulation framework Quantum Espresso [31]. At first, the geometric structure of the trilayer unit cell was relaxed until the force on each atom in each direction was less than 0.01 eV/Å. The self-consistent convergence criterion for energy was kept fixed at $10^{-9}$ Ry. For geometry optimization and energy calculation of the electronic structure, we used scalar relativistic norm-conserving pseudopotential with Perdew–Burke–Ernzerhof (PBE) exchange-correlation functional [34]. The Brillouin Zone (BZ) sampling was done using a Monkhorst–Pack scheme of 24x24x1 points for electronic structure calculation [35]. After structural optimization and ground state energy calculation, we calculated the band structure of the trilayer structure using the bands.x package. The dielectric constant for the trilayer lattice structure was calculated using ph.x package of Quantum Espresso. The results obtained from electronic structure simulation on the trilayer heterostructures under relaxed condition is given in Table I.

### B. Device Structure and Simulation Methodology for MOSFET:

The device structure simulated in this work is a double gate MOSFET as shown in Fig. 1(a) with 10 nm gate length. The 2D ballistic transport simulator has been developed using self-consistent method. At first, 2D Poisson equation is solved to extract the potential and conduction band profile, $E_C(x,z)$ inside the device. After that, an average conduction band profile, $\overline{E_C}(z)$ (equation (1)) is used to solve 1D Schrodinger equation along gate confinement direction (equation (2)).

$$\overline{E_C}(z) = \frac{1}{L_x} \int_0^{L_x} E_C(x,z)dx \tag{1}$$

$$[-\frac{\hbar^2}{2}\frac{\partial}{\partial y}(\frac{1}{m_z^*}\frac{\partial}{\partial z}) + \overline{E_C}(z)]\overline{\psi_{sub}^m}(z) = \overline{E_{sub}^m}.\overline{\psi_{sub}^m}(z) \tag{2}$$

Solving 1D Schrodinger equation gives us the average subband energy, $\overline{E_{sub}^m}$ and wavefunction, $\overline{\psi_{sub}^m}(z)$ for the $m^{th}$ subband. Using First Order Perturbation Theory, the eigen energies in the channel is calculated as shown in equation (3). Then 1D Hamiltonian matrix is formed along the transport direction and the retarded Green's function $G(E)$ is formulated (equation (4)).

Here, $\Sigma_S$ and $\Sigma_D$ are self-energy matrices for source and drain contacts respectively. The self-energy matrices can be further used to evaluate spectral density matrices, $A_S$ and $A_D$ (equation (5))[32].

$$E_{sub}^m(x) = \overline{E_{sub}^m} + \int_z E_C(x,z) |\overline{\psi_{sub}^m}(z)|^2 \, dz - \int_z \overline{E_C}(x,y) |\overline{\psi_{sub}^m}(z)|^2 \, dz \tag{3}$$

$$G(E) = [EI - H - \Sigma_S(E) - \Sigma_D(E)]^{-1} \tag{4}$$

$$A_S = G\Gamma_S G^\dagger \quad \text{and} \quad A_D = G\Gamma_D G^\dagger \tag{5}$$

$$\Gamma_S = i(\Sigma_S - \Sigma_S^\dagger) \quad \text{and} \quad \Gamma_D = i(\Sigma_D - \Sigma_D^\dagger) \tag{6}$$

Here, $\Gamma_S$ and $\Gamma_D$ represents broadening matrices for source and drain contacts respectively. 2-D carrier density for the $m^{th}$ subband, $n_{2D}^m$ in the channel, is calculated using equation (7) [32].

$$n_{2D}^m(E) = \frac{1}{\hbar a}\sqrt{\frac{m_y^* k_B T}{2\pi^3}} \int_{-\infty}^{\infty} (A_S^m \Im_{-1/2}(E, \mu_S) + A_D^m \Im_{-1/2}(E, \mu_D)) dE \tag{7}$$

The calculated charge density is fed back into the Poisson's equation and the self-consistent loop continues. Once self-consistency is achieved, we calculated the drain current for the $m^{th}$ subband, $I_{DS}^m$ using equation (8).

$$I_{DS}^m(E) = \frac{q}{\hbar^2}\sqrt{\frac{m_y^* k_B T}{2\pi^3}} T^m(E)[\Im_{-1/2}(E, \mu_S) - \Im_{-1/2}(E, \mu_D)]$$
(8)

$$T^m(E) = Trace(\Gamma_S(E) G^m(E) \Gamma_D(E) G^{m\dagger}(E)) \tag{9}$$

Here, $T^m(E)$ is the transmission probability calculated using equation (9).

The physics of interband tunneling has been exploited in MOSFETs to design low power switching devices. In MOSFETs, although carrier transport takes place over the potential barrier at the source-channel junction, interband tunneling could still affect device performance in the sub-threshold regime when carrier effective mass and bandgap get lowered in the channel. In this study, we have incorporated the effect of interband tunneling using the formulation depicted in [36].

*C. Device Structure and Simulation Method for Nanobiosensor*

Fig. 1(b) and 1(c) show the schematic of the proposed double gate FET used in this work as pH sensor and biosensor respectively. Trilayer TMDC heterostructure as channel material with thicknesses around 2 nm is used. $HfO_2$ has been used as a gate dielectric on both sides of the channel for this work. However, the simulation procedure used in this work can take into account of various dielectrics. In the case of pH sensing, fluid/front gate voltage, $V_{FG}$ is kept 1V for all simulations while Back gate voltage, $V_{BG}$ is changed from 1V to 5V for operation over the Nernst limit. The thickness of both top gate and bottom gate oxide is varied for pH sensing.

Beside pH sensing, we also discuss the application of the proposed trilayer TMDC as a possible channel material in a potentiometric Nanobiosensor for protein detection. The device prototype in Fig. 1(c) has been inspired from [37] where the original channel material Si is replaced by the TMDC heterostructure. The device is incorporated with proper receptors to provide a more realistic conclusion than the simple approach used in [23].

We have considered an artificial protein structure (Aspartic acid) where amino acids are tagged to a histidine chain. A part of this artificial protein remains uncharged since no amino acids are attached there. By contrast, the rest of the histidine backbone is negatively charged since we consider Aspartic acids that carry one negative charge each for binding to the tag. In this work, the charge of the Aspartic acids has been varied from a single charge up to nine charges. Therefore, for different Aspartic acids, we will get a different surface charge density that will cause a change in sensor response. The electrolyte region includes the histidine-tagged Aspartic acids as well as the neutral part of the tag. The thickness of both top and bottom oxide is kept constant (top oxide 2 nm and bottom oxide 20 nm) to values for which measurable change in device current is found for change in the number of Aspartic acid. The top oxide layer is passivated by an octadecyltrimethoxysilane (ODTMS) monolayer, required for the bio-functionalization of the semiconductor device. Widths of ODTMS, lipid membrane and neutral part of histidine tag have been considered 1.6 nm, 2.0 nm and 2.8 nm respectively. No interface trap is present in top oxide interface because of the functionalization by ODTMS. For this reason, we are not considering any site-binding charges in this case and therefore, no pH sensing is possible with this structure. Lipid membrane has been used as surface functionalization upon ODTMS layer which acts as a receptor for the histidine-tagged Aspartic acid. Material parameters for lipid membrane is kept same as those of ODTMS. Since the lipid membrane layer is highly dense, no electrolyte is present within this layer. We have considered an electrolyte ion concentration of 50mM. For all calculations, the pH of the bulk electrolyte has been set to 7.

Equations governing electrostatics in various regions have been listed in Table II both for pH sensor and potentiometric biosensor. Potential profile along the confinement direction by solving these equations is benchmarked with that of [23] (Not shown). Drain bias is kept to a small value ($V_{ds}$=0.1V) as conventional for this type of sensor and a channel length of 10 µm has been used. Therefore, drift-diffusion model can be reasonably used for current measurement in these cases. For current simulations, we have used D. Jimenez's model [38]. Ohmic contacts are assumed.

◙ *Results and Discussion:*

    A.  *Ballistic Simulation of MoS$_2$/MoSe$_2$/MoS$_2$ MOSFET*

In this section, we would present the ballistic simulation study of the double gate MOSFET with trilayer MoS$_2$/MoSe$_2$/MoS$_2$ heterostructure used as the channel material. For ballistic simulation, only one subband has been considered. In Fig. 2(a), the 1$^{st}$ subband energy and 2D carrier density in the channel at $Vd$=0.5 V when MoS$_2$/MoSe$_2$/MoS$_2$ trilayer is used as the channel material shows that carrier density in the channel increases with increasing gate voltage. Energy resolved current density profile in Fig. 2(b) shows that under the barrier transport is very small compared to over the barrier ballistic transport which refers to strong electrostatic control as channel thickness extremely scaled down. The local density of states characteristics in Fig. 2(c) shows strong interference patterns as carriers are being reflected from the energy barrier at source and drain ends. Fig. 2(d) shows the Id-Vg characteristics at two different drain bias voltage, $Vd$=0.1 V and $Vd$=0.5 V. From the Id-Vg characteristics at $Vd$=0.5 V, the extracted values of sub-threshold swing (SS) and drain induced barrier lowering (DIBL) were 80.256 mV/dec and 26.2 mV/V respectively.

*B. Comparison of Ballistic Simulation Using Different TMDC Heterostructures*

In this section, we will present a comparison of the ballistic performance of double gate MOSFETs simulated using different TMDC trilayers. Fig. 3(a) shows the first subband energy obtained from NEGF simulation for three different heterostructures. For $MoS_2/WS_2/MoS_2$ trilayer, the simulation shows slightly lowered energy barrier height at the source end. On the other hand, the insertion of $MoSe_2$ or $WSe_2$ monolayer between $MoS_2$ monolayers does not cause a significant change in subband profile. This can be attributed to similar carrier effective mass at conduction band minima for these two trilayer heterostructures in AAA stacking under relaxed condition. 2D carrier density in the channel at $Vd$=0.5 V and $Vg$=0.4 V for the heterostructures in Fig. 3(b) shows that with the insertion of a $WS_2$ monolayer, the carrier density in the channel increases slightly. Although $WSe_2$ monolayer and $MoSe_2$ monolayer insertion result in similar subband energy profiles, the insertion of $MoSe_2$ monolayer was seen to be providing a slightly higher carrier density in the channel. Fig. 3(c) shows the gate capacitance of the trilayer heterostructure devices used in this study. As shown in [39], the gate capacitance in a device like MOSFET can be presented as a series combination of two capacitances- the oxide capacitance that depends on gate dielectric and device dimensions and semiconductor capacitance, which is a function of quantum capacitance and change in quantum well shape with applied gate bias. The quantum capacitance is directly proportional to the carrier effective mass. Among the structures studied, according to first principle simulation on electronic properties, the $MoS_2/WS_2/MoS_2$ trilayer provided the highest effective mass and therefore higher quantum capacitance. Therefore, device with $MoS_2/WS_2/MoS_2$ heterostructure configuration gives slightly higher gate capacitance which allows more carrier accumulation in the channel. Fig. 3(d) shows the Id-Vg characteristics of the device at $Vd$=0.5 V in both linear and log scale. Higher carrier density in the channel results in slightly higher drain current for $MoS_2/WS_2/MoS_2$ trilayer device. From the ballistic simulation, we have extracted values of SS and DIBL for different trilayer material systems. The extracted values of SS, DIBL is given in Table III.

*C. The Effect of Tensile Strain on Ballistic Device Performance*

We also explored and observed the effect of bi-axial tensile strain on the ballistic device performance. In TMDC materials, application of bi-axial tensile strain causes semiconductor to metal transition, direct to indirect transition in bandgap nature [34]. In our observation, tensile strain application lowered the bandgap for all the hetero-trilayer structures under study and lowered electron effective mass at conduction band minima. The Id-Vg characteristics in Fig. 4 show that tensile strain application lowers drain current for all the trilayer material systems.

This can be attributed to lower carrier accumulation in the channel at increased tensile strain. Although we see a lowering in drain current, there appears to be no significant change in the sub-threshold characteristics of the device i.e. value of sub-threshold swing (SS) with applied tensile strain in case of $MoS_2/MoSe_2/MoS_2$ and $MoS_2/WS_2/MoS_2$ trilayers. However, for $MoS_2/WSe_2/MoS_2$ trilayer, lower bandgap and carrier effective mass results in an increase in the band to band tunneling current which becomes significant at low gate bias voltage condition and therefore we observe an increase in SS and deterioration in sub-threshold device performance.

### D. *Effect of Underlap Geometry*

We have studied the effect of gate underlap geometry on device performance of trilayer TMDC MOSFETs. Implementation of gate underlap geometry, increases effective channel length which increases series resistance in the channel. Underlap geometry also reduces effective coupling of source and drain contact with the channel and therefore could improve device electrostatic performance by reducing SS. In the study of gate underlap geometry, gate length has been kept fixed at 10 nm and $HfO_2$ has been used as the gate dielectric material.

Fig. 5(a) shows the schematic diagram of underlap geometry. We performed ballistic simulation with gate underlap length varying uniformly from 0 nm to 3 nm in 1 nm step. The other physical device parameters have been kept unchanged. Fig. 5(b) shows the Id-Vg characteristics when $MoS_2/MoSe_2/MoS_2$ trilayer has been used as the channel material. The figure shows a reduction in 'on' current when gate underlap length is increased. This can be attributed to the fact that, increasing channel underlap length increases effective channel length and therefore induces additional resistance in the channel. Fig. 5(c) and 5(d) show the Id-Vg characteristics at different gate underlap lengths for $MoS_2/WSe_2/MoS_2$ and $MoS_2/WS_2/MoS_2$ respectively. However, despite this reduction in 'on' current, increasing gate underlap region length does improve sub-threshold performance by lowering effect of the fringing electric field from source and drain. Proper optimization of gate underlap region therefore, is required to design TMDC transistor that could meet the low sub-threshold requirement for future technology nodes without significantly compromising the 'on' current requirement.

Fig. 6(a) shows the variation of 'on' current calculated at $Vg$=0.4 V and $Vd$=0.5 V at different gate underlap lengths. As can be seen from the figure, at different underlap lengths, the $MoS_2/MoSe_2/MoS_2$ and $MoS_2/WSe_2/MoS_2$ trilayers give almost equal 'on' currents. However, $MoS_2/WS_2/MoS_2$ trilayer improves the 'on' current performance of the device. Fig. 6(b) shows the variation of sub-threshold swing with gate underlap length. In all cases, we see a decrease in SS with increased gate underlap length. Of these material systems, MoS$_2$/WSe$_2$/MoS$_2$ trilayer shows highest sub-threshold swing values which can be attributed to its low bandgap.

### E. Applications as Sensors

In this paper, for the pH sensor, we have varied top gate oxide thickness, $T_{TOP}$ and bottom gate oxide thickness, $T_{BOT}$ for three different TMDC heterostructure FETs separately to find out how sensitivity changes with scaling and material parameter. In DGFET pH sensors, one sweeps the bottom gate (BG) bias, instead of the fluid gate (FG), to obtain the transfer characteristics ($I_d$-$V_{BG}$) whereas a fixed bias is maintained at the fluid gate, and the corresponding pH sensitivity is measured in terms of the threshold voltage shift. Due to asymmetry of top and bottom oxide thickness, the resultant asymmetry in top and bottom oxide capacitances originates the high pH sensitivity [41] of this sensor according to the following equation:

$$\frac{\Delta V_{BG}}{\Delta pH} = \alpha_{SN} \left(\frac{C_{tox}}{C_{box}}\right)\left(\frac{\Delta V_{FG}}{\Delta pH}\right) \quad (10)$$

In this work, we have used high gate bias for the front gate ($V_{FG}$ = 1V). So, $\alpha_{SN}$ will be close to one [24][41][42] for this work. That is why back gate threshold voltage will vary approximately linearly with the change of pH considering that $\left(\frac{\Delta V_{FG}}{\Delta pH}\right)$ will be less than the Nernst limit and be almost constant during the sweep of back gate voltage. Approximately identical super-Nernst sensitivity is obtained for all trilayer FETs for a wide range of operation [pH 4 to 8] for various back oxide thicknesses as seen from Fig 7(a). Another point to note from Fig 7(a) is that sensitivity increases almost linearly with the increase of back oxide thickness while keeping front oxide thickness fixed at 2 nm. However, it must be mentioned that the sensitivity reported in this work will be an upper level estimation of not yet experimentally measured sensitivity because of the assumption made in section II (C).

The increase of spread of drain current for pH 4 to 8 in the sub-threshold region with the increase of back oxide thickness results in a shift in threshold voltage, $\Delta V_{T, BG}$ as shown in Fig. 7(b). It is evident from the equation 10 and Fig 7(c) that increasing the top oxide thickness results in a reduction of sensitivity while the opposite trend is observed for bottom oxide. As evident from equation 10, an increase in $T_{TOP}$ will reduce top oxide capacitances, $C_{tox}$ ultimately decreasing the sensitivity. This finding is also consistent with the trend found in literature [24].

Since the trend is similar for all three trilayer FETs, we have shown output for only $MoS_2/WSe_2/MoS_2$ DGFET pH sensor. To evaluate the prospect of these materials in a FET-based Nano biosensor, we have considered a more realistic structure of Fig. 1(c). We have varied the no. of Aspartic acid charges to find out the sensitivity of these sensors. Sensitivity in the case of the biosensor is defined as the ratio of the difference in current before and after biomolecule binding to the lower of the two currents [27]. The magnitude of the negative protein charge density increases with the number of Aspartic acids. This results in a lower potential in the charged part of the protein region. As a result, the surface potential (potential at top gate oxide-receptor interface) decreases with increasing protein charge. This potential acts as top gate voltage in the current simulator. Since the surface potential is decreasing with the increasing no. of Aspartic acid, current will decrease for all three devices as seen from Fig. 8.

From Fig. 9, which is obtained from Fig. 8, it is noticeable that, all three FETs show highest sensitivity (approximately $10^5$ or above) in the sub-threshold region. Among all these three FETs, $MoS_2/WS_2/MoS_2$ FET shows the lowest sensitivity for a wide region of operation. The observed trend of high sensitivity in sub-threshold region is quite similar to that of experimentally reported Silicon FET [43] as well as multilayer $MoS_2$ based biosensors [26][27].

As seen from the Fig. 9, the relative change in transistor 'on' current with the increasing no of Aspartic acid is relatively small compared to that in sub-threshold regime. This is because the FET is already conducting a high current in 'on' condition, so a small change in surface potential due to the attachment of a biomolecule results in a corresponding small change in the drain current. However, in the completely off device or in sub-threshold regime, FET conducts little or no current. Therefore, a small change in surface potential due to binding of protein brings relatively larger change in drain current. This phenomenon can be explained from another viewpoint. In the sub-threshold region, the drain current has an exponential dependence on the gate dielectric surface potential, while in saturation and linear regions the relationship is quadratic and linear, respectively. Hence, the sensitivity in the sub-threshold region is much higher compared to those in the saturation and linear regions. These findings indicate that biosensor operation in sub-threshold regime will optimize the sensor response for these heterostructure FETs while improving the lower limit of bio molecule detection at the same time.

◉ *Conclusion*

In this work, we have performed a ballistic simulation study on the performance of van der Waal trilayer TMDC heterostructures as channel materials in MOSFETs for sub-10 nm region operation. Of the three trilayers studied, $MoS_2/WS_2/MoS_2$ trilayer appears to be showing better sub-threshold performance and higher drain current due to higher carrier accumulation. $MoS_2/WSe_2/MoS_2$ trilayer shows degraded sub-threshold performance due to its low bandgap and lower carrier effective mass. Application of tensile strain causes a reduction in drain current for all the materials studied. Implementation of gate underlap geometry could improve sub-threshold performance at the cost of 'on' current reduction of these devices. We have also investigated the application of TMDC trilayer heterostructures as a channel material in pH and potentiometric biosensor. According to our study, pH sensors incorporating these materials could show above Nernst sensitivity because of the bottom gate operation in a DGFET structure. These materials can be considered as viable options in implementing FET-based biosensor because of their high sensitivities especially in the sub-threshold regime.

◉ *Reference:*